\documentclass[aps,nobibnotes,preprint,groupedaddress,showkeys,showpacs]{revtex4}
\usepackage{graphicx}
\usepackage{dcolumn}
\usepackage{amsmath}
\usepackage{amssymb}
\usepackage{bm}

\begin{document}
\input{epsf}

\preprint{APS/123-QED}

\title{Computer Simulations of Pulsatile Human Blood Flow
  Through 3D-Models of the Human Aortic Arch, Vessels of Simple Geometry
  and a Bifurcated Artery: Investigation of Blood Viscosity and
  Turbulent Effects}


\author{Renat A. Sultanov}
\email{rasultanov@stcloudstate.edu;r.sultanov2@yahoo.com}
\affiliation{Business Computing Research Laboratory, St. Cloud State University,
St. Cloud, MN 56301, USA}
\author{Dennis Guster}
\email{dcguster@stcloudstate.edu}
\affiliation{Business Computer Information Systems, St. Cloud State University,
St. Cloud, MN 56301, USA}

\date{\today}

\begin{abstract}
We report computational results of blood flow through a model of
the human aortic arch and a vessel of actual diameter and length.
On the top of the aortic arch the branching of the 
arteries are included: the subclavian and jugular.
A realistic pulsatile flow is used in all simulations.
Calculations for bifurcation type vessels are also carried out and presented.
Different mathematical methods for numerical solution of the
fluid dynamics equations have been considered. The non-Newtonian behaviour of
the human blood is investigated together with turbulence effects.
A detailed time-dependent mathematical convergence test has been carried out.
The results of computer simulations of the blood flow in vessels of three
different geometries are
presented: for pressure, strain rate and velocity component
distributions we found significant disagreements between our results obtained 
with realistic non-Newtonian treatment of human blood and the widely
used method in the literature: a simple Newtonian approximation. 
A significant increase of the strain rate and, as a result,
a wall shear stress distribution, is found in the region of the aortic arch. 
Turbulent effects are found to be important, particularly in the case of bifurcation vessels.
\end{abstract}



\maketitle
\section{Introduction}
Cardiovascular diseases, such as ischemic heart disease, myocardial
infarction, and stroke
are leading causes of death in Western countries. All of these 
vascular diseases share 
a common element: atherosclerosis. They also share a common final event: the 
failure or destruction of the vascular wall structure \cite{1-lee05,2-dhein05}.

Atherosclerosis reduces arterial lumen size through plaque formation and arterial 
wall thickening and it occurs at specific arterial sites. 
This phenomenon is related to hemodynamics and to wall shear stress
(WSS) distribution \cite{fung93}. From the physical point of view
WSS is the tangential drag force produced by moving blood, and
it is a mathematical function of the velocity gradient of blood near
the endothelial surface \cite{landau}. Arterial wall remodeling is
regulated by WSS \cite{grotberg04}.
For example, in response to high shear stress arteries enlarge. 

Further, from the biomechanical point of view one can conclude, that
the atherosclerotic plaques localize preferentially in the regions of
low shear stresses, but not in regions of higher shear stresses.
Furthermore, decreased shear stress induces intimal thickening in vessels which
have adapted to high flow.

Final vascular events that induce fatal outcomes, such as 
acute coronary syndrome, are triggered by the sudden mechanical 
disruption of an arterial 
wall. Thus, we can conclude, that the final consequences of tragic
fatal vascular diseases 
are strongly connected to mechanical events that occur on the vascular
wall, and these, in turn, which are likely to be heavily influenced by
alterations in blood flow and the characteristics of the blood itself. 

Currently researchers in the field of biomechanics and biomedicine conduct
laboratory investigations of human blood flow in different shape and size
tubes, which are designed to be approximate models of human vessels
and arteries, see for example \cite{19,2}. 
Some researchers also carry out intensive computer simulations of
these bio-mechanical systems, see for example
\cite{362,1991,9,chn04,8,morris2004,chn06,dur2007,pre08,renat08,renat08a}. 

Also, in some special laboratory works specific stents are 
incorporated in such artifical vessels (tubes). Stent implantation has
been used to open diseased coronary blood vessels, allowing improved
perfusion of the cardiac muscle. Used in combination with drug therapy,
vascular repair and dilation techniques (angioplasty) the use of
metallic stents has created a multibillion dollar industry.
Stents are commonly used in many different blood vessels, but the primary
site of deployment is in diseased coronary arteries.

Stents represent a very special case in the
modeling research problems mentioned above \cite{77}.
Taking into account that stents have a very
small size and rather complicated structure and shape, this situation makes it
difficult to obtain precise measurements. Therefore high quality and
precise computer simulations of blood flow through vessels with
implanted stents would be most useful \cite{77}.
Work of this type is already underway and
we would like to mention several pertinent studies \cite{6a,3a,10a,11a}.

\vspace{7mm}

Nevertheless, there are still many difficulties
in obtaining precise realistic geometries for the required vessels.
Human arteries, especially the aorta, have complicated 
spatial-geometric and characteristic configurations. For example, the aortic 
arch centerline does not lie
on a plane and there are major branches at the top of the arch
feeding the carotid arterial circulation. One of the main problems in
the field of bio-medical blood flow simulation is to obtain precise
geometrical-mathematical representations of different vessels. This
information in turn needs to be included in the simulation programs.

\vspace{7mm}

However, in our opinion, as a first step of these investigations
it would be useful to apply simpler 3D-geometry forms and models,
but to take into account as much as possible the precise physical effects of blood movement
such as the non-Newtonian characteristics of
human blood, realistic pulsatile flow, and possible turbulent effects.
Because of pulsatility the last one may be significant in
contributing to the final results of this study.

\vspace{7mm}

In the current work we carried out real-time full-dimensional
computer simulations of a realistic pulsatile human blood
flow in actual size vessels, vessels with a bifurcation, and in a
model of the aortic arch.
We take into account different physical effects, such as turbulence
and the non-Newtonian nature of human blood.
The next section presents the mathematical methodology and the
physical model used in this
work. The general purpose commercial computational fluid dynamics
program FLOW3D is used for its basic functionality, but we supplemented
its capability by adding our routines to obtain the results presented in this work.

Sec. 3 presents results for three vessels of different geometries.
The {\sf CGS} unit system is used in
all simulations, as well as for presentation of the
results. Conclusions and discussion comparing our results to well
respected studies is included in Sec. 4.

\section{Mathematical methodology and physical models}

As we mentioned above, we undertook pulsatile human blood flow
simulation experiments using different size and shape human vessel arteries.
For each spatial configuration one needs to provide a
specific approach for the numerical solution to the complicated second
order partial differential equations of the fluid dynamics (FD).
For example, for simple cylindrical vessels we used the cylindrical
coordinate system: $\vec r = {(r, \theta, \phi)}$.
However, for the aortic arch or bifurcated vessels, where there is no
cylindrical symmetry, we
applied the Cartezian coordinate system:  $\vec r = {(x, y, z)}$. In the
cases of the aortic arch and bifurcated vessels we used up to five
blocks of matched Cartezian coordinate subsystems.

Below we present the FD equations in a general form,
because, for each of the special cases, considered in this work
and the chosen coordinate system, the partial differential
equations of the fluid dynamics may look different, although we
understand, that the general
differential operator form of the equations is always unique.

\subsection{Equations}

The equations of motion for the fluid velocity components $(u, v, w)$
in the 3-coordinate system are the Navier-Stokes equations
with specific additional terms included in the FLOW3D program:
\begin{eqnarray}
\frac{\partial u}{\partial t}+\frac{1}{V_F}
(
{uA_x\frac{\partial u}{\partial x}}+
{vA_yR\frac{\partial u}{\partial y}}+
{wA_z\frac{\partial u}{\partial z}}
)
-\xi \frac{A_y v^2}{x V_f} = 
-\frac{1}{\rho}\frac{\partial p}{\partial x}+
G_x+f_x-b_x- \nonumber \\  
\frac{R_{sor}}{\rho V_f}(u-u_w-\delta \cdot u_s)
\label{eq:2a}
\end{eqnarray}
\begin{eqnarray}
\frac{\partial v}{\partial t}+\frac{1}{V_F}
(
{uA_x\frac{\partial v}{\partial x}}+
{vA_yR\frac{\partial v}{\partial y}}+
{wA_z\frac{\partial v}{\partial z}}
)
+\xi \frac{A_y uv}{x V_f} = 
-\frac{R}{\rho}\frac{\partial p}{\partial y}+
G_y+f_y-b_y- \nonumber \\
\frac{R_{sor}}{\rho V_f}(v-v_w-\delta \cdot v_s)
\label{eq:2b}
\end{eqnarray}
\begin{eqnarray}
\frac{\partial w}{\partial t}+\frac{1}{V_F}
(
{uA_x\frac{\partial w}{\partial x}}+
{vA_yR\frac{\partial w}{\partial y}}+
{wA_z\frac{\partial w}{\partial z}}) = 
-\frac{1}{\rho}\frac{\partial p}{\partial z}+
G_z+f_z-b_z- \nonumber \\
\frac{R_{sor}}{\rho V_f}(w-w_w-\delta \cdot w_s).
\label{eq:2c}
\end{eqnarray}
Here, $(u, v, w)$ are the velocity components in coordinate
directions $(x,y,z)$ respectively. For example,
when Cartesian coordinates are used, $R=1$ and $\xi=0$, 
see FLOW3D manual \cite{flow3d}. $A_x$ is the fractional area
open to flow in the $x$ direction, analogously for $A_y$ and $A_z$.
Next,
$V_F$ is the fractional volume open to flow, 
$R$ and $\xi$ are coefficients which depend on the coordinate system:
$(x,y,z)$ or $(r,\theta,z)$, $\rho$ is the fluid density,
$R_{sor}$ is a mass source term.
Finally, $(G_x, G_y, G_z)$ are so called body accelerations \cite{flow3d},  $(f_x, f_y, f_z)$ are
viscous accelerations,  $(b_x, b_y, b_z)$ are the flow losses in porous
media or across porous baffle plates, and the final term accounts for
the injection of mass at a source represented by a geometry component.
Mathematical expressions for the viscous accelerations  $(f_x, f_y,
f_z)$ are presented in the Appendix.


The term $U_w=(u_w,v_w,w_w)$ in equation (\ref{eq:2a})
is the velocity of the source component, which will generally be
non-zero for a mass source at a General Moving Object (GMO) \cite{flow3d}.
The term $U_s=(u_s,v_s,w_s)$ is the velocity of the fluid at the surface 
of the source relative to the source itself. It is
computed in each control volume as 
\begin{equation}
\vec U_s=\frac{1}{\rho_s}\frac{d(Q \vec n)}{dA}
\end{equation}
%
%
%
where $dQ$ is the mass flow rate, $\rho_s$ fluid source density, 
$dA$ the area of the source surface in the cell
and $\vec n$ the outward normal to the surface. 
The source is of the stagnation pressure type
when in equations. (\ref{eq:2a}-\ref{eq:2c}) $\delta=0.0$. Next,
$\delta=1.0$ corresponds to the source of the static pressure type.


As we already mentioned, in all simulations we considered the blood flow
as a pulsatile flow. The final result for the inflow waveform has been 
taken from figure 3 of work \cite{greece2007}. The values of the velocity waveform,
which were used in our simulations,
are shown in Table 1 depending on pulse time. The pulse was applied for
5.5 cycle times in our work.  

These velocity values are used  as
time-dependent inflow initial boundary conditions. These numbers are included
directly in the FLOW3D program.

Next, the general mass continuity
equation, which is solved within the FLOW3D program has the following
{\it general} form:
\begin{eqnarray}
V_f\frac{\partial \rho}{\partial t} + 
\frac{\partial}{\partial x}(\rho u A_x)+
R\frac{\partial}{\partial y}(\rho v A_y)+
\frac{\partial}{\partial z}(\rho w A_z)+  
\xi \frac{\rho u A_x}{x} = R_{dif}+R_{sor},
\end{eqnarray}
$R_{dif}$ is a turbulent diffusion term, and $R_{sor}$ is a mass source.
The turbulent diffusion term is
\begin{eqnarray}
R_{dif}=
\frac{\partial}{\partial x}(v_pA_x\frac{\partial \rho}{\partial x})+R
\frac{\partial}{\partial y}(v_pA_yR\frac{\partial \rho}{\partial
  y})+   
\frac{\partial}{\partial z}(v_pA_z\frac{\partial \rho}{\partial z})+
\xi\frac{\rho v_pA_x}{x},
\end{eqnarray}
where the coefficient $v_p=C_p\mu / \rho$, $\mu$ is dynamic viscosity and
$C_p$ is a constant. The $R_{sor}$ term is a density source term that
can be used to model mass injections through porous obstacle surfaces.

Compressible flow problems require the solution of the
full density transport equation In this work we treated blood as
an incompressible fluid. For incompressible fluids, $\rho=constant$ and
the equation (2) becomes the following:
\begin{equation}
\frac{\partial}{\partial x}(u A_x)+
\frac{\partial}{\partial y}(v A_y)+
\frac{\partial}{\partial z}(w A_z)+
\xi \frac{u A_x}{x} = \frac{R_{sor}}{\rho}.
\end{equation}

It is assumed, that at a stagnation pressure source fluid
enters the domain at zero velocity. As a result, pressure
should be considered at the source to move the fluid away from 
the source. For example, such sources are designed to model fluid
emerging at the end of a rocket or the simple deflating process of a balloon. 

In general, stagnation pressure sources apply to cases
when the momentum of the emerging fluid is created 
inside the source component, like in a rocket engine.

At a static pressure source the fluid velocity is 
computed from the mass flow rate and the surface area of the
source. In this case, no extra pressure is required 
to propel the fluid away from the source. An example of
such a source is fluid emerging from a long straight pipe. 
Note that in this case the fluid momentum is created
far from where the source is located.

Turbulence models can be taken into account in FLOW3D. It allows us to
estimate the influence of turbulent fluctuations on mean flow
quantities. This influence is usually expressed by additional
diffusion terms in the equations for mean mass, momentum, and
energy. The turbulence kinetic energy per unit mass, $q$, is the
following
\begin{eqnarray}
\frac{\partial q}{\partial t}+\frac{1}{V_F}
\left (
uA_x\frac{\partial q}{\partial x}+
vA_yR\frac{\partial q}{\partial y}+
wA_z\frac{\partial q}{\partial z}
\right ) =  
P+G+Diff-D,
\end{eqnarray}
where $P$ is shear production, $G$ is buoyancy production, $Diff$
is diffusion, and $D$ is a coefficient \cite{flow3d}.

When the turbulence option is used, the viscosity is a sum of the
molecular and turbulent values. For non-Newtonian 
fluids the viscosity can be a function of the 
strain rate and/or temperature. A general expression
based on the Carreau model is used in FLOW-3D
for the strain rate dependent viscosity:
\begin{eqnarray}
\mu = \mu_{\infty}+\frac{\mu_0E_T-\mu_{\infty}}{\lambda_{00}+
[\lambda_0+(\lambda_1E_T)^2e_{ij}e_{ij}]^{(1-n)/2}}           
+ \frac{\lambda_2}{\surd (e_{ij}e_{ij})},
\label{eq:mu1}
\end{eqnarray}
where
%
$
e_{ij}=1/2(\partial u_i/\partial x_j +\partial u_j/\partial x_i)
$
%
is the fluid strain rate in Cartesian tensor notations, $\mu_{\infty},
\mu_0, \lambda_0, \lambda_1, \lambda_2$ and $n$ are constants. Also,
$
E_T = exp[a(T^*/(T-b) - C)],
$
%
where $T^*, a, b,$ and $c$ are also parameters of the temperature
dependence, and $T$ is fluid temperature. This basic formula is used
in our simulations for blood flow in vessels and in the aortic arch.
For a variable dynamic viscosity $\mu$, the viscous accelerations have
a special form. It is shown in the Appendix.


The equations of fluid dynamics should be solved together with
specific boundary conditions. 
The numerical model starts with a computational mesh, or grid. It consists of a 
number of interconnected elements, or 3D-cells. These 3D-cells subdivide the physical space into 
small volumes with several nodes associated with each such volume.
The nodes are used to store values of the 
unknown parameters, such as pressure, strain rate,
temperature, velocity components and etcetera. 
This procedure provides values for defining the flow
parameters at discrete locations and allows specific boundary
conditions to be set up.


\section{Numerical results}

Results of our simulations are presented below.
One of the most important preliminary testing tasks
is to check for numerical convergence. This test has been
successfully accomplished in this work. A portion of the test calculation results are
shown bellow in this paper.
Next, in this work particular attention has been given to
the calculations of wall shear stress distribution (WSS).
WSS is the tangential drag force produced by moving blood.
It is a mathematical function of the velocity gradient of blood near
the endothelial surface:
%
$
\tau_w=\mu \left [\partial U(t,y,R_v) / \partial y \right ]_{y \approx 0}.
$
%
Here $\mu$ is the dynamic viscosity, $t$ is current time, $U(t,y,R_v)$
is the flow velocity parallel to the wall, $y$ is the distance to
the wall of the vessel, and $R_v$ is its radius.
It was shown, that the magnitude of WSS
is directly proportional to blood flow/blood viscosity and
inversely proportional to the cube of the radius of the vessel,
in other words a small change of the radius of a vessel will have a large 
effect on WSS.




\subsection{Straight vessel: cylinder}

First, we chose a simple vessel geometry,
that is we considered the shape of a straight vessel to be a tube. 
In our simulations involving a straight cylinder type vessel
we applied a cylindrical coordinate system: $(r, \theta, Z)$
with the axis $OZ$ directed over the tube axis.
Different quantities of cells 
have been used to discretize the empty
space inside the tube. In the open space (inner part of
the tube) the fluid dynamics equations have been solved using appropriate
mathematical boundary conditions.
The size
of the tube is: $L=8$ cm (in length) and $R=0.34$ cm (length of inner radius). The thickness
of the vessel wall is $s=0.03$ cm. We have applied 5.5 cycles of blood pulse.

Let us now consider in more detail the expression (\ref{eq:mu1}). In these calculations
we follow the work \cite{1991},   
where the Carreau model of the
human blood has also been used. In consistence with \cite{1991}    
we choose the following coefficients:
$\lambda_2=\lambda_{00}=0$, $a=0$ and $E_T=1$, that is we don't take into
account the temperature dependence of the viscosity. This investigation
can be done in our subsequent work. Next: $\lambda_0=1$,
$\lambda_1=3.313$ sec, $\mu_{\infty}=0.0345$ P, $\mu_0=0.56$ P, and $n=0.3568$.


Time-dependent results for pressure, strain rate and 
velocity components $V$ and $W$ are presented in
Fig. 1. The turbulent effects are not taken into account.
We chose to present only one precise geometrical point for
comparison purposes:
the middle point: $r=\theta=0$, and $Z=4.0$ cm.
%
%
%
The data for Fig. 1. were obtained with the non-Newtonian model of human blood.
We refer the reader to the comments provided for the figure. We were able
to closely replicate the values for all previous cell sizes \cite{flow3d}
and obtain almost identical values, for example for pressure, wall
shear stress and other parameters, for 0.065 and 0.062 cell sizes \cite{flow3d}.
This means, that the convergence has been achieved.

Next, it would be very interesting to compare the results calculated with and without 
the turbulent effect. To support this endeavor we use the realistic
non-Newtonian model of blood viscosity, the pulsatile flow, and
the size of computation cells at which convergence has been achieved,
that is the 0.062 size for all computational cells \cite{flow3d}.

The results are presented in Fig. 2. As we see from Fig. 2 the effect of the
turbulence is significant, particularly in regard to dynamic viscosity and
strain rate. This result means that in the case of pulsatile flows and
non-Newtonian viscosity the turbulent term should be taken into account.
In Fig. 3 we separately show the results for the pulsatile pressure distribution
and the turbulent energy, again using the middle point of the cylinder.
Finally, it would also be very interesting to make a comparison
between the
results calculated using both a Newtonian and non-Newtonian
viscosity. However, as in previous simulations, we will apply the pulsatile
flow with the turbulence included, since it has proved to be important.
The results are shown in Fig. 4. As one can see, we obtain
significant differences between these two calculations. 
We specifically observed that
for the pressure distribution, dynamic viscosity and turbulent
energy, we obtained significant disagreements.

Thus, we arrive at the important conclusion: within a time-dependent
(pulsatile) flow of human blood it is necessary to take into
account turbulence and non-Newtonian viscosity.

\subsection{Hemodynamics in the coronary bifurcation}

Below we show the result of our subsequent simulation involving a 90${^{\circ}}$
bifurcated coronary artery in Figs. 5 and 6.
The geometrical model of the bifurcation consisted of a 90${^{\circ}}$
intersection of two cylinders. This model represents the
bifurcation between the left anterior descending coronary artery and
the circumflex coronary artery.
In our opinion, in the case of pulsatile flow it is
more interesting to present results in
a time-dependent way. This method can provide a wider picture of highly
non-stationary flow systems. In this paper, because of space
limitations, we just included time-dependent results for pressure,
dynamic viscosity, turbulent energy, and strain rate. However, we
understand, that results which depend on spatial
coordinates $(r, \theta, Z)$ for a few fixed moments of time are also highly useful.

In the case of the bifurcation shown in Fig. 5, we report the results for only two
spatial points, which are the two outflow sides: the far right side and the
farthest upper side of the bifurcation. The length of the lower horizontal
vessel is 4 cm and its diameter is 0.54 cm. The length of the upper vertical
vessel is 1.2 cm and its diameter is 0.4 cm. These sizes are
consistent with average size human vessels.

In Fig. 5 blood flows in from the left to the right with the imposed initial
velocity profile taken from Table 1. The pressure, strain rate and turbulent
energy distributions are shown for only one specific time moment
$t$=4.329 s. The velocity vectors are also shown on these plots.

Further, Fig. 6 represents our time-dependent results for the two outflow sides
mentioned above. These results are for pressure, dynamic
viscosity, turbulent energy and strain rate. The bold black lines
are the results for the right outflow side, and the red dashed lines are
the results for the farthest upper side (see comments to Fig. 6).
In conclusion,
the main goal of these calculations is to adopt them to investigate a
case in which a stent \cite{77} is implanted in the bifurcation area. 


\subsection{Blood flow in aortic arch}

The geometry of the blood simulations inside the human aortic arch is
shown in Fig. 7. On the top of the aortic arch three arteries
are included. These arteries deliver the blood to the carotid artery
and then to the brain. This configuration only models and approximately
represents the real aortic arch. One of the goals of our simulations is to
reveal the physics of the blood flow dynamics in this important portion of the
human cardiovascular system.

The aortic arch is represented as a curved tube.   
The outer radius of the tube is 2.6 cm. A straight vessel (tube) is
also merged to the arch. The length of the straight tube is about 4 cm. Again,
the thickness of the wall is 0.03 cm, and the inner radius of the tube
is $r=0.34$ cm. Once again we are using the Cartesian coordinate system. We also
carried out a convergence test. To better represent the shape of the arch
we applied five Cartesian sub-coordinate systems in our FLOW3D simulations.
After the discretization the total number of all cubic cells reached about 900,000.
It is important to mention here, that we again obtained full
numerical convergence.
In this work we computed pressure,
velocity and strain rate distributions in the arch, while the human
blood is treated as a non-Newtonian liquid    
and while the realistic pulsatile blood flow is used.  

In Fig. 7 we present the results of strain rate distributions inside the arch
for two specific time moments. At the
most left point, which is the inlet, we specify the pulsatile velocity
source as the initial condition, that is the data from table 1 are used. From
the general theory of fluid mechanics \cite{landau} it is possible to determine
together with the blood density and viscosity, and spatial geometries,
the dynamics of the blood according to the Navier-Stokes equation and
its boundary conditions.
Small vectors indicate the blood velocity. As can be seen from Fig. 7 blood flows
from left to right in direction. However, because of pulsatility
blood flows in the opposite direction too. 

The values of the strain
rate are also shown. These values are strongly oscillating. From
the plots one can conclude that in the region of the arch the strain
rate values become much larger than in the region of the straight vessel.
This result represents clear evidence that in this
part of the human vascular system atherosclerotic plaques should localize less than in
the straight vessels.
However, the higher wall shear stress values in the aortic arch could be
the reason for sudden mechanical disruption of the arterial wall in this
part of the human vascular system. These results are consistent
with laboratory and clinical observations. In Fig. 8 we depict the
pressure distribution in the arch.

\section{Conclusion}

In this work we applied computational fluid dynamics techniques to support
pulsatile human blood flow simulations through different shape/size vessels and the
aortic arch. The realistic blood pulse has been adopted and applied from work \cite{greece2007}.
The geometrical size of the vessels and the aortic arch have been selected to match
the average real values. Human blood was treated in two different ways: (a) as a
Newtonian liquid when the viscosity of the blood has a constant
value, and (b) as a non-Newtonian liquid with the viscosity value
represented by the equation (\ref{eq:mu1}). The numerical coefficients in
(\ref{eq:mu1}) have been taken from work \cite{1991}.

It is always difficult to obtain a steady-state cycle profile and
stable computational results at the very beginning of time-dependent simulations.
However, after a short stabilization period a steady-state cycle profile can be obtained.
In our simulations we used up to 5.5 pulse cycles to reach complete
steady state profiles. We obtained valid results for pressure,
wall shear stress distribution and other physical parameters, such as
the three velocity components of blood flow. All of
these were shown in Figs. 1 and 2.

Our simulations showed that the FLOW3D program
is capable of providing stable numerical results for all geometries included in this work.
The time-dependent mathematical convergence test has been successfully carried out.
Particular attention has been paid to this aspect of the calculations.
It is a well known fact that fluid dynamics equations can
have unstable solutions \cite{landau}.  
Therefore, numerical convergence has been tested and confirmed in this work.

The result of computer simulations of blood flow in vessels for three
different geometries have been presented. For pressure, strain rate and velocity component
distributions we found significant disagreements between our results obtained 
with the realistic non-Newtonian treatment of human blood and the widely
used method in literature: a simple Newtonian approximation. 

Our results are in good agreement with the conclusions of works
\cite{chn04,chn06}, where the authors also obtained significant differences between their
results calculated with and without the non-Newtonian effect of blood viscosity.
However, the recent work \cite{austr07} should be mentioned, in which
the authors performed 2-dimensional simulations of human
blood flow through the carotid artery with and without the non-Newtonian
effect of the viscosity. 
They did not find any substantial differences in their results. 
But in the paper \cite{india08}, where the authors also performed
simulations for the carotid artery, only the non-Newtonian viscosity
was used.

Next, the influence of a possible turbulent effect has also been investigated in
this work. It was found that the effect is important. We believe, that the
physical reason of this phenomena lies in the strong pulsatility of
the flow and in 
the non-Newtonian viscosity of the blood. The contribution of the turbulence is most
significant in the area of bifurcated vessels.

Finally, a significant increase of the strain rate and,  
the wall shear stress distribution, is found in the region of the aortic arch. 
This computational result provides additional evidence to support
recent clinical and laboratory observations that this part of the human
cardiovascular system is under higher risk of disruption \cite{carter2001,poch2006}.
In future works it would be interesting to include the elasticity of the walls of
the aortic arch \cite{pre94} and other vessels.

In conclusion, we would like to specifically point out, that the developments in
this work can be directly applied to even more interesting and very
important situations such as when a stent is implanted inside
a vessel \cite{77}. In this case, for example, it would be very useful to determine
blood flow disturbance, the pressure distribution, strain rate and
values of other physical parameters. The results of this work should
allow us to determine the optimal size and shape of effective stents.
As we mentioned in the Introduction some
research groups are carrying out laboratory and computer simulations of blood flow
through vessels with implanted stents \cite{77}. It is very
difficult to underestimate the value of these works.


\section{Appendix}

For a variable dynamic viscosity $\mu$, the viscous accelerations are
\begin{eqnarray}
\rho V_F f_x = w^s_x -
[
\frac{\partial}{\partial x}(A_x\tau_{xx}) +
R\frac{\partial}{\partial y}(A_y\tau_{xy}) + 
\frac{\partial}{\partial z}(A_z\tau_{xz}) + 
\frac{\xi}{x}(A_x\tau_{xx}-A_y\tau_{yy})
]
\label{eq:wss7}
\end{eqnarray}
\begin{eqnarray}
\rho V_F f_y = w^s_y - [
\frac{\partial}{\partial x}(A_x\tau_{xy}) +
R\frac{\partial}{\partial y}(A_y\tau_{yy}) + 
\frac{\partial}{\partial z}(A_z\tau_{yz}) + 
\frac{\xi}{x}(A_x+A_y\tau_{xy})
]
\end{eqnarray}
\begin{eqnarray}
\rho V_F f_z = w^s_z - [
\frac{\partial}{\partial x}(A_x\tau_{xz})+
R\frac{\partial}{\partial y}(A_y\tau_{yz}) + 
\frac{\partial}{\partial z}(A_z\tau_{zz}) + 
\frac{\xi}{x}(A_x\tau_{xz})],
\label{eq:wss77}
\end{eqnarray}
\vspace{5mm}\\
where
{\small
\begin{eqnarray}
\tau_{xx}=-2\mu\left(\frac{\partial u}{\partial
x}-\frac{1}{3}\left(\frac{\partial u}{\partial x}+
R\frac{\partial v}{\partial y}+
\frac{\partial w}{\partial z}+
\frac{\xi u}{x}\right)\right)
\label{eq:wss1}
\end{eqnarray}
\begin{eqnarray}
\tau_{yy}=-2\mu \left[ R\frac{\partial v}{\partial
x}+\xi\frac{u}{x} - 
\frac{1}{3}\left(\frac{\partial u}{\partial x}+
R\frac{\partial v}{\partial y}+
\frac{\partial w}{\partial z}+
\frac{\xi u}{x}\right ) \right ]
\end{eqnarray}
\begin{eqnarray}
\tau_{zz}=-2\mu\left(\frac{\partial w}{\partial
z}-\frac{1}{3}\left(\frac{\partial u}{\partial x}+
R\frac{\partial v}{\partial y}+
\frac{\partial w}{\partial z}+
\frac{\xi u}{x}\right)\right)
\end{eqnarray}
\begin{eqnarray}
\tau_{xy}=-\mu\left(\frac{\partial v}{\partial x}+
R\frac{\partial u}{\partial y}-\frac{\xi v}{x}\right)
\end{eqnarray}
\begin{eqnarray}
\tau_{xz}=-\mu\left(\frac{\partial u}{\partial z}+
\frac{\partial w}{\partial x}\right)
\end{eqnarray}
\begin{eqnarray}
\tau_{yz}=-\mu\left(\frac{\partial v}{\partial z}+
R\frac{\partial w}{\partial y}\right).
\label{eq:wss6}
\end{eqnarray}
}
In the above equations (\ref{eq:wss7})-(\ref{eq:wss77})
the terms $w^s_x, w^s_y$ and $w^s_z$ are wall shear stresses.
If these terms are equal to zero, there is no wall shear stress. This is
because the remaining terms contain the
fractional flow areas $(A_x, A_y, A_z)$ which vanish
at the walls \cite{flow3d}.




\newpage
\begin{table} 
\caption{Time-dependent velocity profile at the vessel inlet.
The velocity vector is directed exactly over the axis of the tube.
Blood velocities $v_i$ are given in cm/sec versus the corresponding time moments $t_i$
in sec ($i=1,2,...,30$). Results taken from Fig. 2 of work \cite{greece2007}.}
\begin{ruledtabular}
\begin{tabular}{cccccc}
$t_i$  &  $v_i$   &    $t_i$   &  $v_i$   &   $t_i$    &  $v_i$  \\
0.0   & 1.5 &  0.35  & 27.5 &  0.625 & -4.0\\
0.015 & 1.6 &  0.385 & 47.0 &  0.65  & -3.0\\
0.035 & 2.5 &  0.39  & 49.0 &  0.7   & 2.1 \\
0.075 & 2.0 &  0.4   & 50.0 &  0.725 & 2.6 \\
0.1   & 2.6 &  0.41  & 49.0 &  0.75  & 2.1 \\
0.15  & 2.0 &  0.415 & 47.0 &  0.8   & 2.6 \\
0.2   & 2.6 &  0.45  & 33.0 &  0.85  & 3.2 \\
0.23  & 2.2 &  0.5   & 14.5 &  0.9   & 2.6 \\
0.25  & 3.6 &  0.55  & 1.9  &  0.95  & 2.2 \\
0.3   & 12.0&  0.4   & 50.0 &  1.0   & 1.5 \\
\end{tabular}
\end{ruledtabular}
\end{table}

\clearpage
\begin{center}
\underline{\bf Figure Captions}
\end{center}

{\bf FIG. 1.}
Test of numerical convergence.
Time-dependent dynamic viscosity, strain rate and velocity
components V and W. Results for a vessel of
simple geometry - cylinder type,
for a specific spatial point inside the cylinder - the middle point.
No turbulence effects are involved in these simulations with the realistic
non-Newtonian viscosity of human blood.
Black dashed line: calculations with 0.08 size for all cells
\cite{flow3d}, red dot-dashed line
with 0.07, green double dot - dashed line with 0.065, and
blue bold line calculations with 0.062 size for all cells.\\

{\bf FIG. 2.}
Time-dependent results for a specific geometrical point inside the cylinder: the
middle point. Black dashed line: simulations without taking into
account the turbulence;
red bold line results with the turbulence. The non-Newtonian viscosity 
is taken into account.\\

{\bf FIG. 3.}
Time-dependent results for pressure and the turbulent
energy in the middle point of the cylinder. The non-Newtonian viscosity.\\

{\bf FIG. 4.}
Results for pressure, dynamic viscosity, turbulent energy and
velocity W. Time-dependent results for the middle point of the
cylinder. Bold black line calculations with non-Newtonian viscosity
of the human blood; red dashed line with its Newtonian approximation.\\

{\bf FIG. 5.}
Time-dependent results for a vessel with bifurcation. Pulsatile blood flow,
non-Newtonian viscosity, and the turbulence effect is included.
Bold black line: results for the far right outflow side $z=0.0$; red
dashed line results for the farthest up outflow side $y=0.0$.\\

{\bf FIG. 6.}
The figures are 2D-plots showing
the blood flow in the bifurcated vessels for only one precise moment of the
discretized time  $t_i=4.329$ sec, the corresponding index is $i=40$.
Upper plot represents the result for the pressure distribution in the
bifurcation, and
the pressure ranges from 2068 dynes/sq-cm to 6758 dynes/sq-cm.
The middle plot represents the results for the strain rate distribution and
the lower plot shows results for the turbulent energy in the bifurcation.
The range of the values is also shown.\\

{\bf FIG. 7.}
Blood flow in the aortic arch. These two plots represent the full
2D-picture of the geometry used in these simulations.
Shaded results for the strain rate are also shown, the bars
on the right show the values.
Results are for two specific moments of the time
$t_{40}$ = 4.329 sec and $t_{41}$ = 4.440 sec.
The values of the strain rate distribution range 
from 0.0 1/sec to 357.0 1/sec (upper plot) and from  0.0 to 671 1/sec
(lower plot).
The maximum values of the strain rate are localized in the region inside
the arch. Blood flows from right to left in both pictures.\\

{\bf FIG. 8.}
These two plots represent in more detail
the region of the arch together with shaded results for the pressure distribution.
The bars on the right show the values.
These results are for two specific moments of the time 
$t_{40}$ = 4.329 sec and $t_{41}$ = 4.440 sec, where the pressure ranges
from 957 dynes/sp-cm to 5889 dynes/sq-cm (upper plot), and from -616 dynes/sq-cm
to 7070 dynes/sq-cm (lower plot).\\

\newpage
\begin{center}
\underline{\bf Figures}
\end{center}

\begin{figure}[ht]
\includegraphics*[scale=1.0,width=27pc,height=27pc]{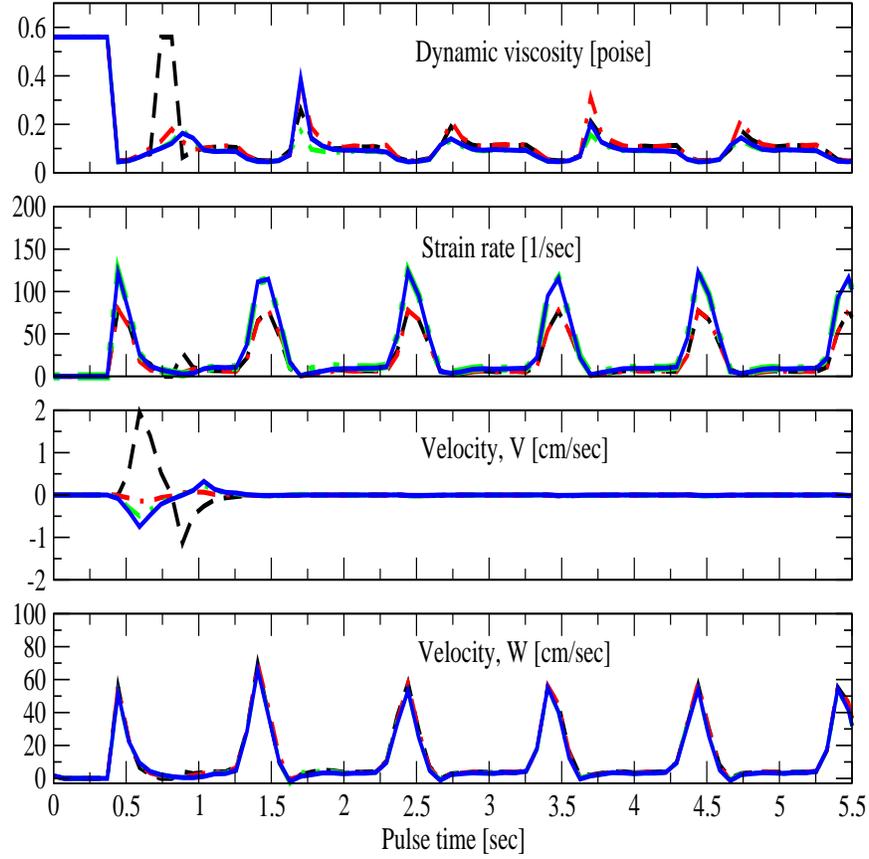}
\caption{
Test of numerical convergence.
Time-dependent dynamic viscosity, strain rate and velocity
components V and W. Results for a vessel of
simple geometry - cylinder type,
for a specific spatial point inside the cylinder - the middle point.
No turbulence effects are involved in these simulations with the realistic
non-Newtonian viscosity of human blood.
Black dashed line: calculations with 0.08 size for all cells
\cite{flow3d}, red dot-dashed line
with 0.07, green double dot - dashed line with 0.065, and
blue bold line calculations with 0.062 size for all cells.}
\label{fig:fig1}
\end{figure}

\newpage
\begin{figure}[ht]
\includegraphics*[scale=1.0,width=27pc,height=27pc]{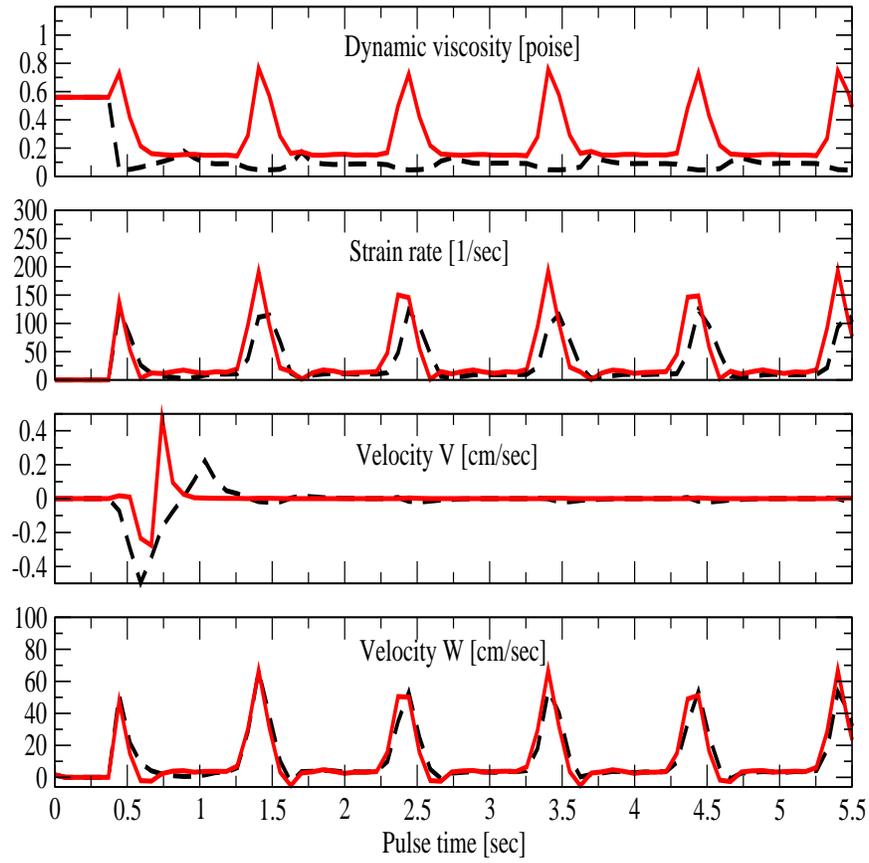}
\caption{Time-dependent results for a specific geometrical point inside the cylinder: the
middle point. Black dashed line: simulations without taking into
account the turbulence;
red bold line results with the turbulence. The non-Newtonian viscosity is taken into account.}
\label{fig:fig2}
\end{figure}

\newpage
\begin{figure}[ht]
\includegraphics*[scale=1.0,width=27pc,height=27pc]{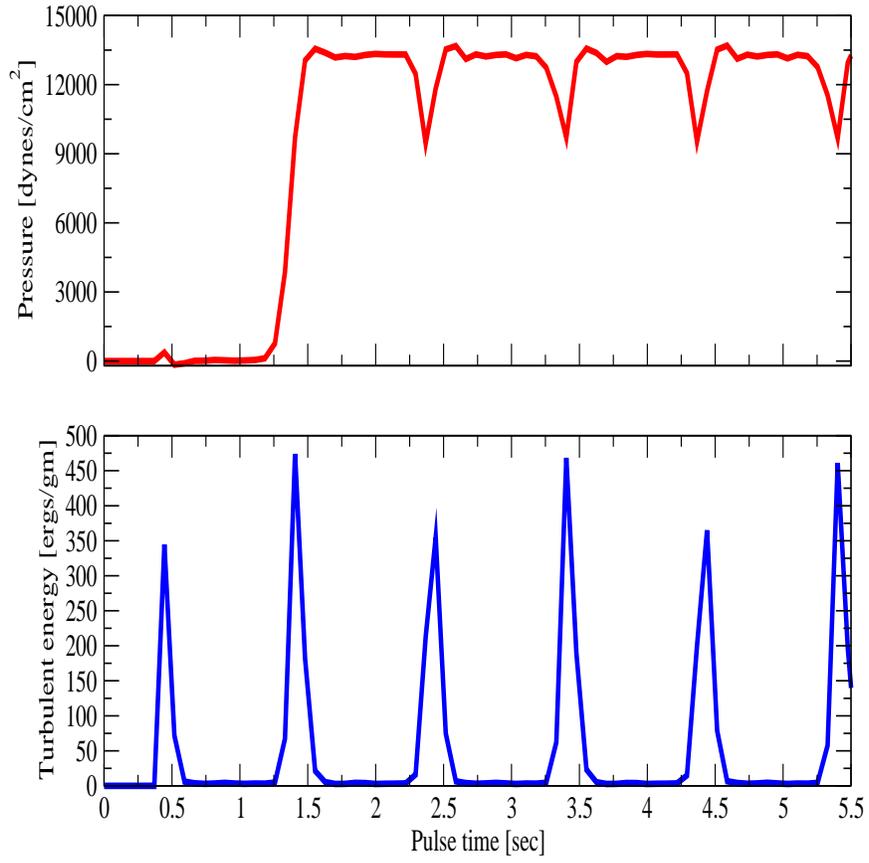}
\caption{Time-dependent results for pressure and the turbulent
  energy in the middle point of the cylinder. The non-Newtonian viscosity.
}
\label{fig:fig3}
\end{figure}

\newpage
\begin{figure}[ht]
\includegraphics*[scale=1.0,width=27pc,height=27pc]{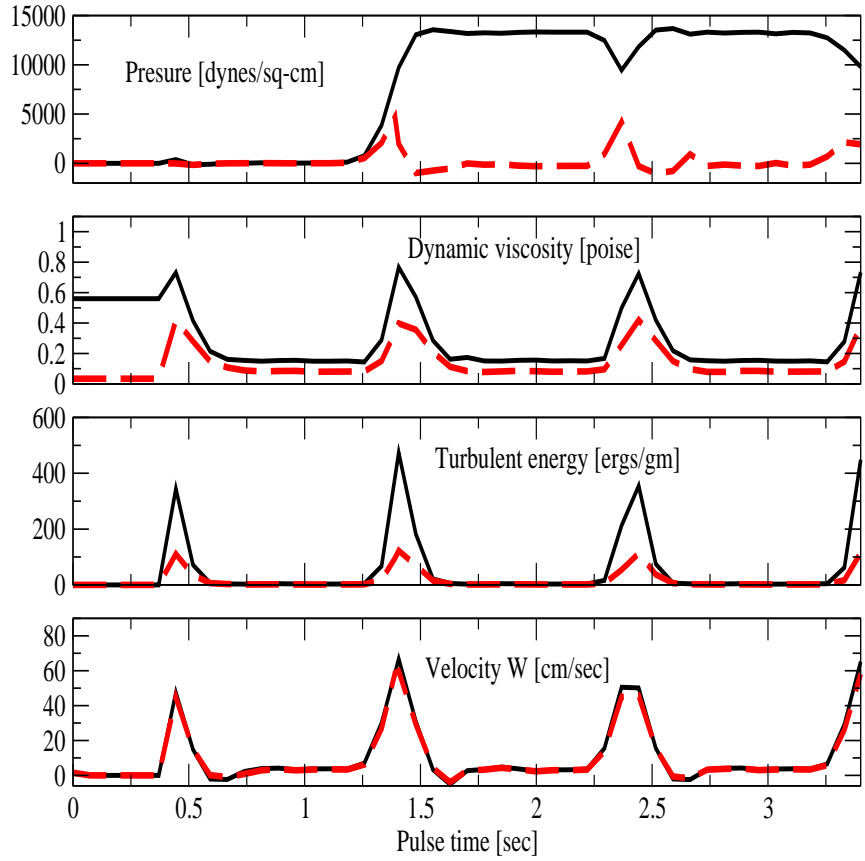}
\caption{Results for pressure, dynamic viscosity, turbulent energy and
  velocity W. Time-dependent results for the middle point of the
  cylinder. Bold black line calculations with non-Newtonian viscosity
  of the human blood; red dashed line with its Newtonian approximation.
}
\label{fig:fig4}
\end{figure}

\newpage
\begin{figure}[ht]
\includegraphics*[scale=1.0,width=27pc,height=27pc]{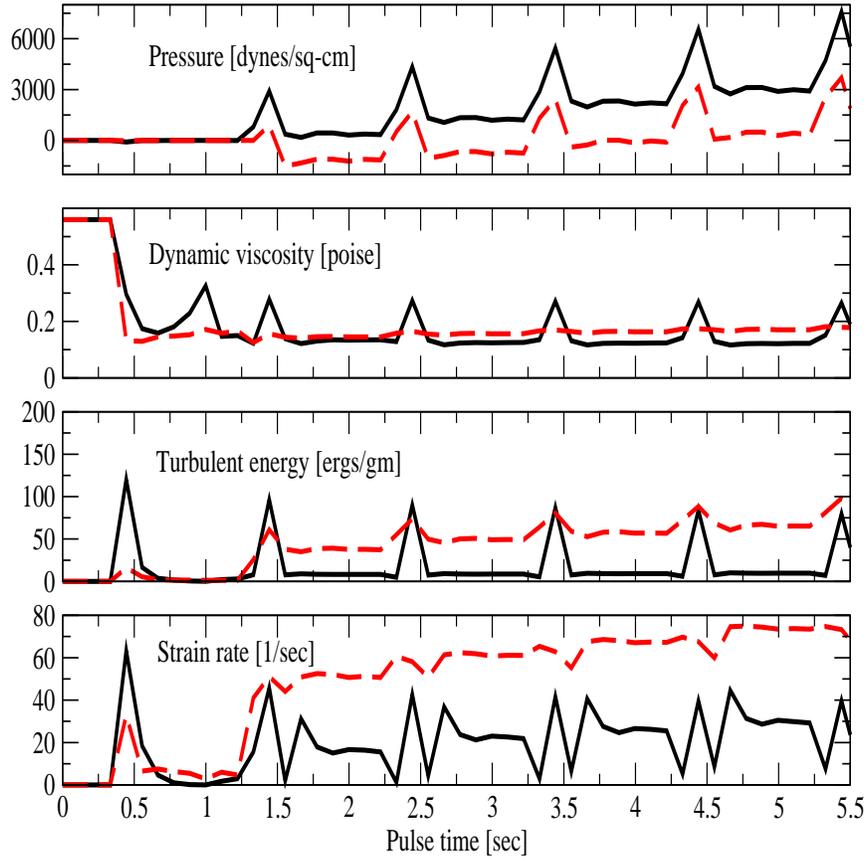}
\caption{Time-dependent results for a vessel with bifurcation. Pulsatile blood flow,
non-Newtonian viscosity, and the turbulence effect is included.
Bold black line: results for the far right outflow side $z=0.0$; red
dashed line results for the farthest up outflow side $y=0.0$
}
\label{fig:fig5}
\end{figure}

\newpage
\begin{figure}[ht]
\vspace{-4.0cm}
\begin{center}
\includegraphics[scale=1.0,width=23pc,height=23pc]{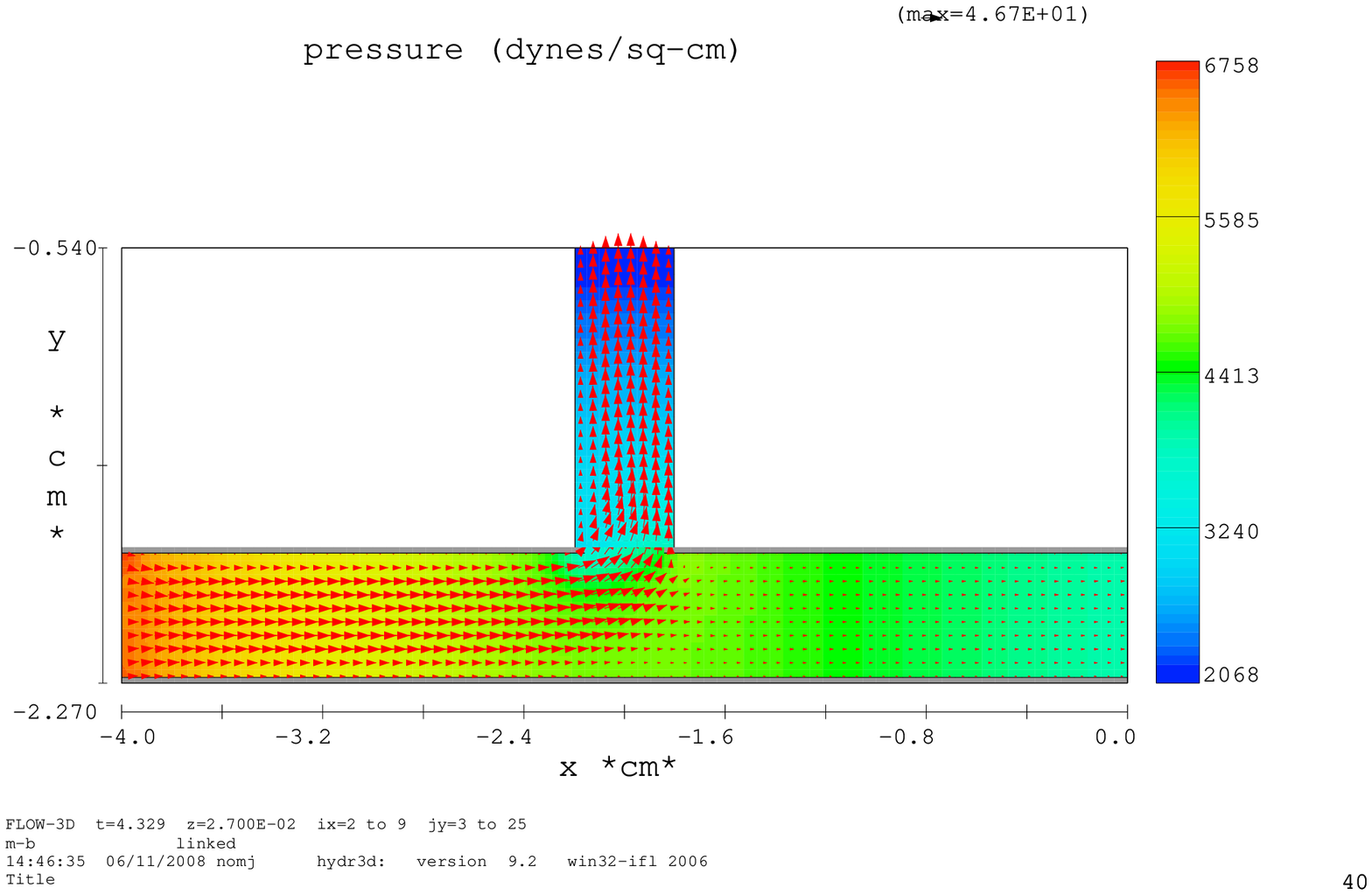}\\
\vspace{-3.0cm}
\includegraphics[scale=1.0,width=23pc,height=23pc]{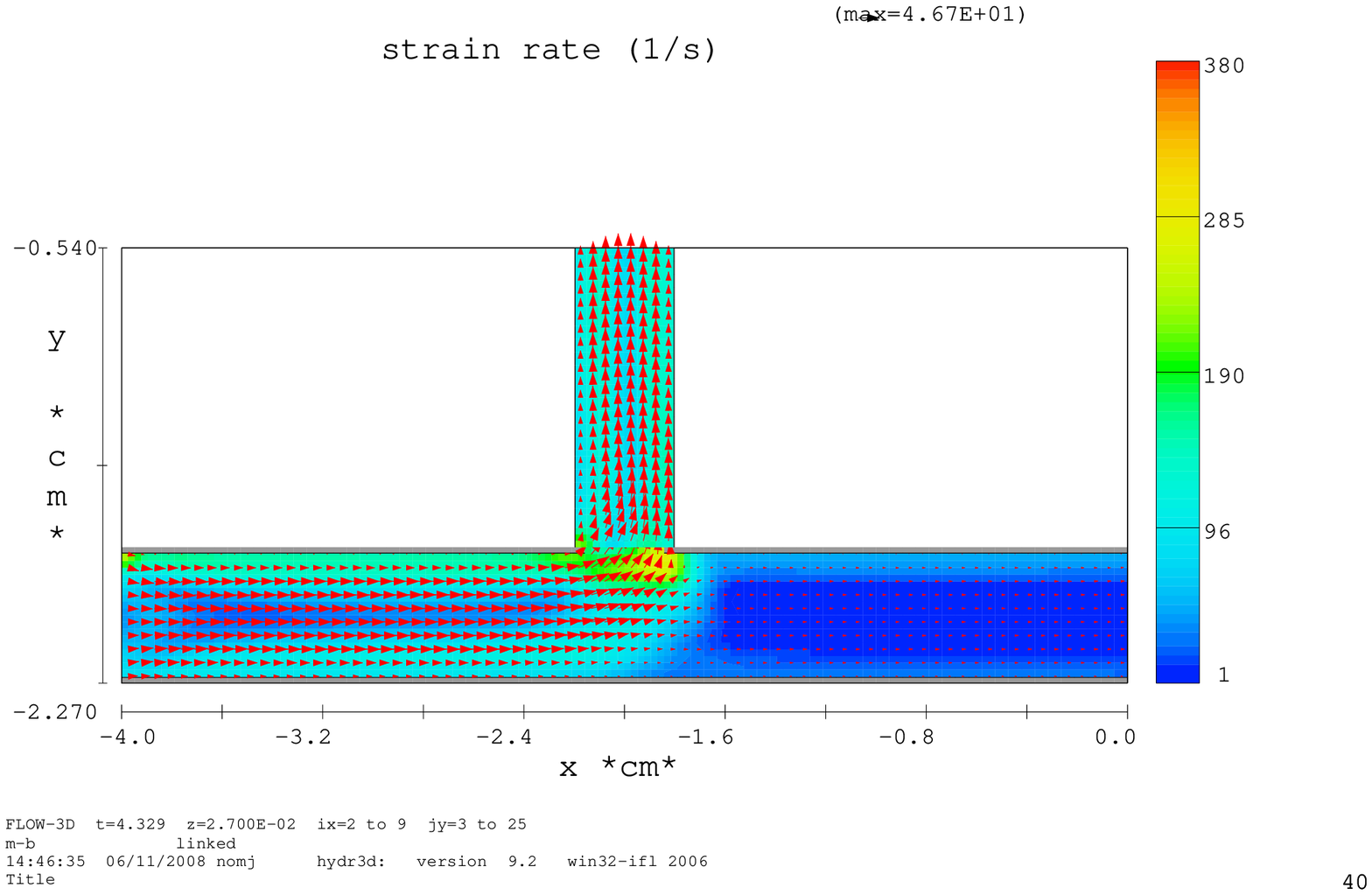}\\
\vspace{-3.0cm}
\includegraphics[scale=1.0,width=23pc,height=23pc]{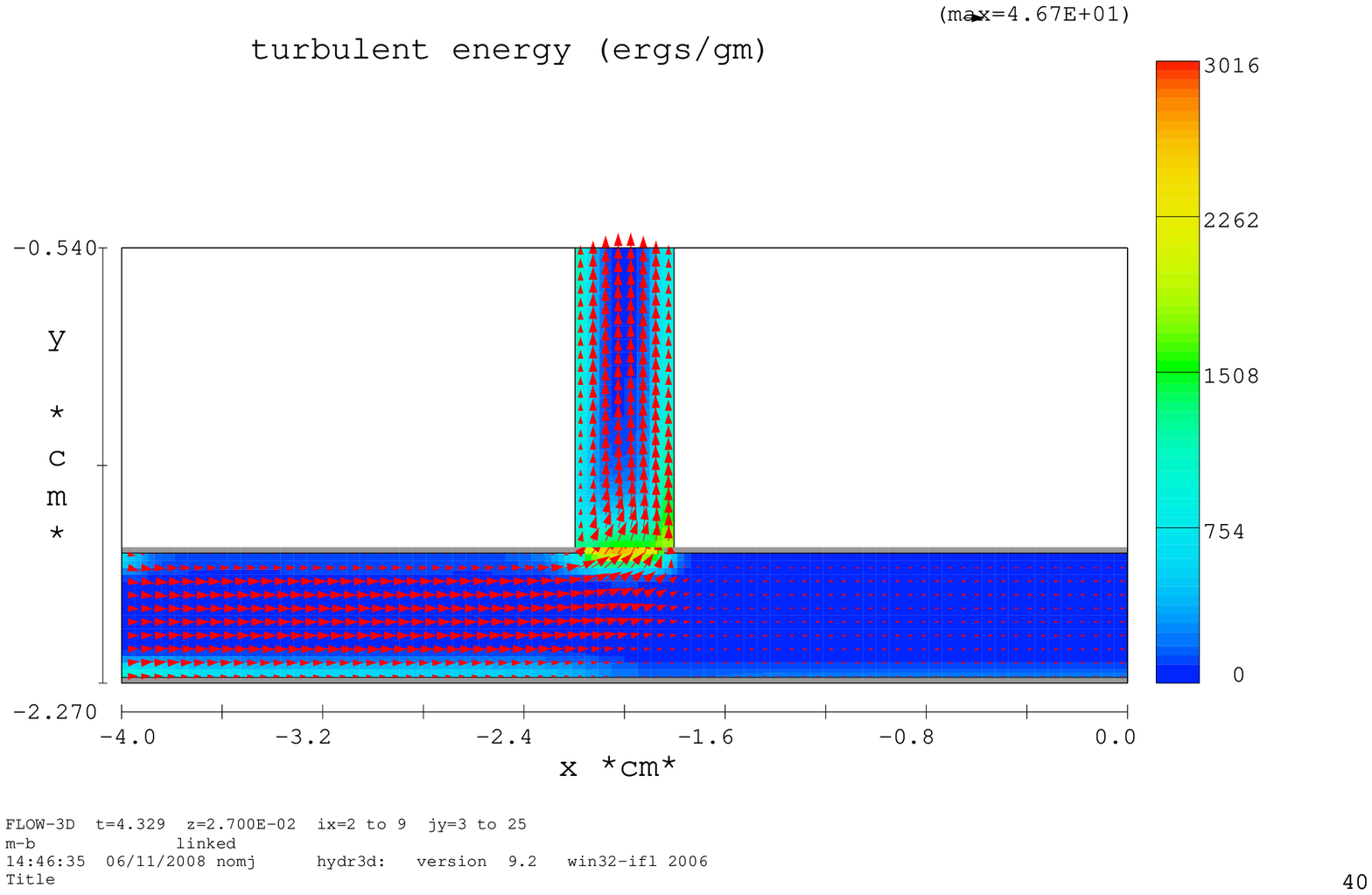}\\
\end{center}
\vspace{-1.2cm}
\caption{
The figures are 2D-plots showing
the blood flow in the bifurcated vessels for only one precise moment of the
discretized time  $t_i=4.329$ sec, the corresponding index is $i=40$.
Upper plot represents the result for the pressure distribution in the
bifurcation, and
the pressure ranges from 2068 dynes/sq-cm to 6758 dynes/sq-cm.
The middle plot represents the results for the strain rate distribution and
the lower plot shows results for the turbulent energy in the bifurcation.
The range of the values is also shown.
}
\label{fig:fig6}
\end{figure}

\newpage
\begin{figure}[ht]
\vspace{-3.3cm}
\begin{center}
\includegraphics[scale=1.0,width=25pc,height=25pc]{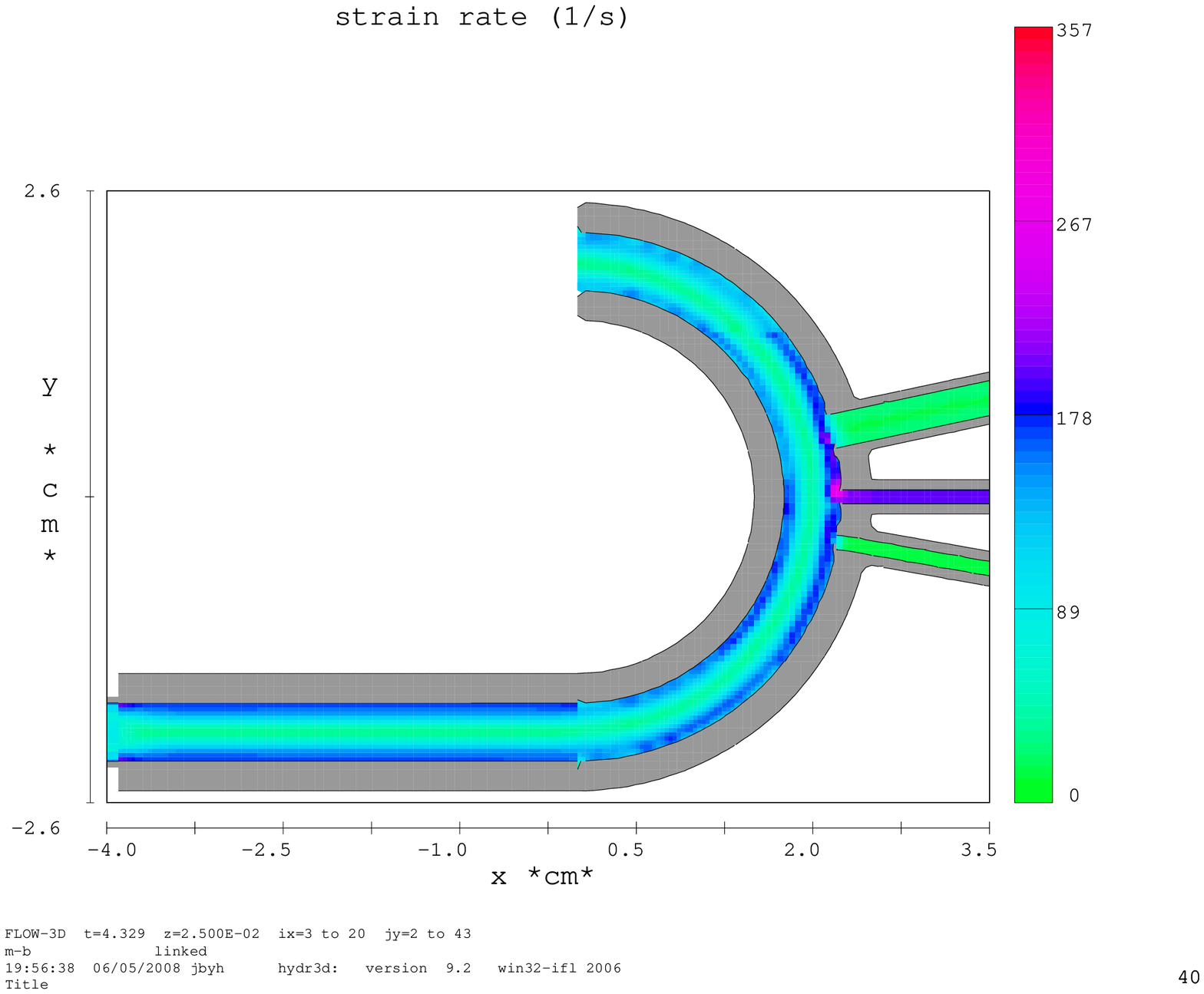}
\includegraphics[scale=1.0,width=25pc,height=25pc]{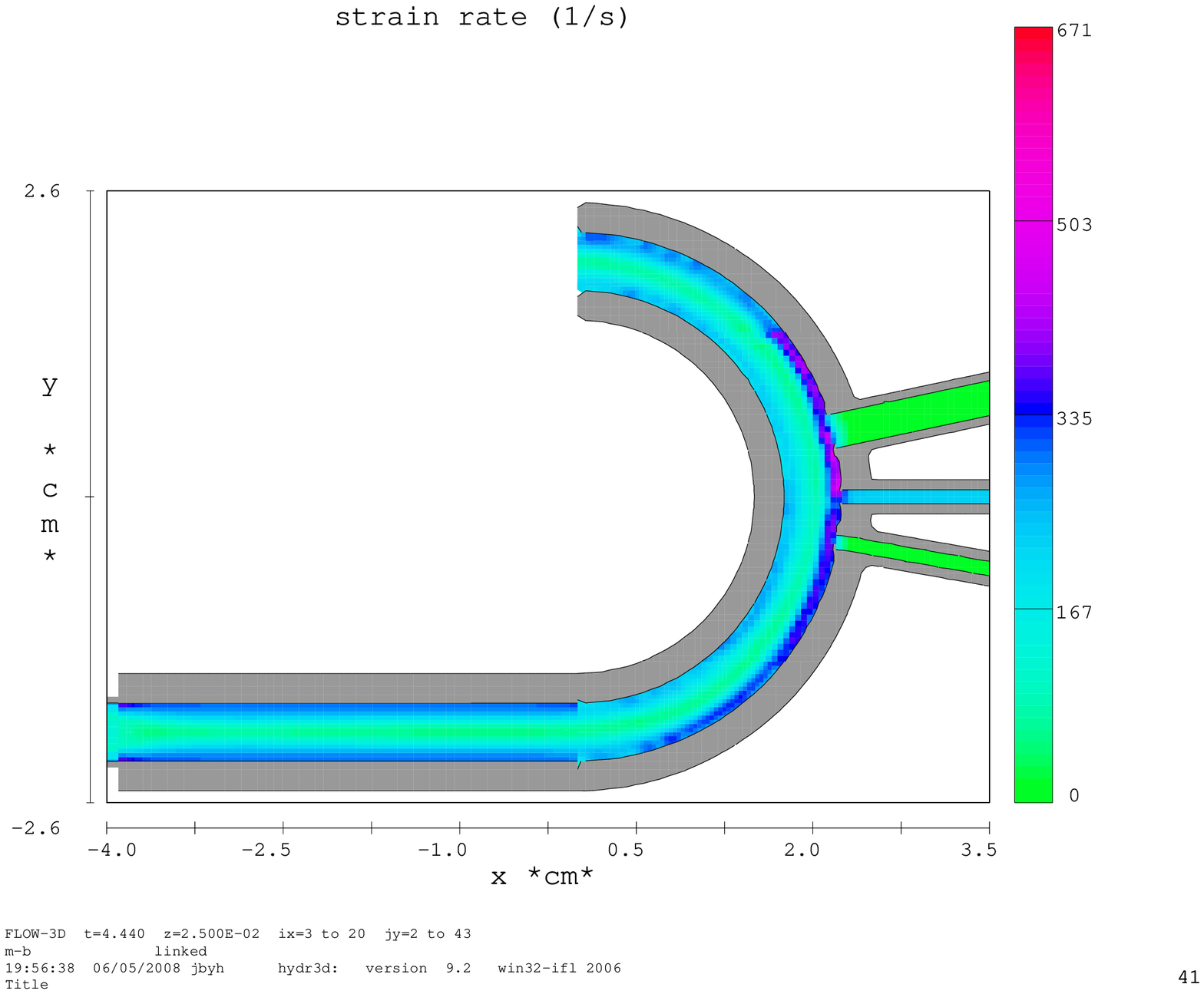}
\end{center}
\caption{Blood flow in the aortic arch. These two plots represent the full
2D-picture of the geometry used in these simulations.
Shaded results for the strain rate are also shown, the bars
on the right show the values.
Results are for two specific moments of the time
$t_{40}$ = 4.329 sec and $t_{41}$ = 4.440 sec.
The values of the strain rate distribution range 
from 0.0 1/sec to 357.0 1/sec (upper plot) and from  0.0 to 671 1/sec
(lower plot).
The maximum values of the strain rate are localized in the region inside
the arch. Blood flows from right to left in both pictures.}
\label{fig:fig7}
\end{figure}

\newpage
\begin{figure}[ht]
\vspace{-3.3cm}
\begin{center}
\includegraphics[scale=1.0,width=25pc,height=25pc]{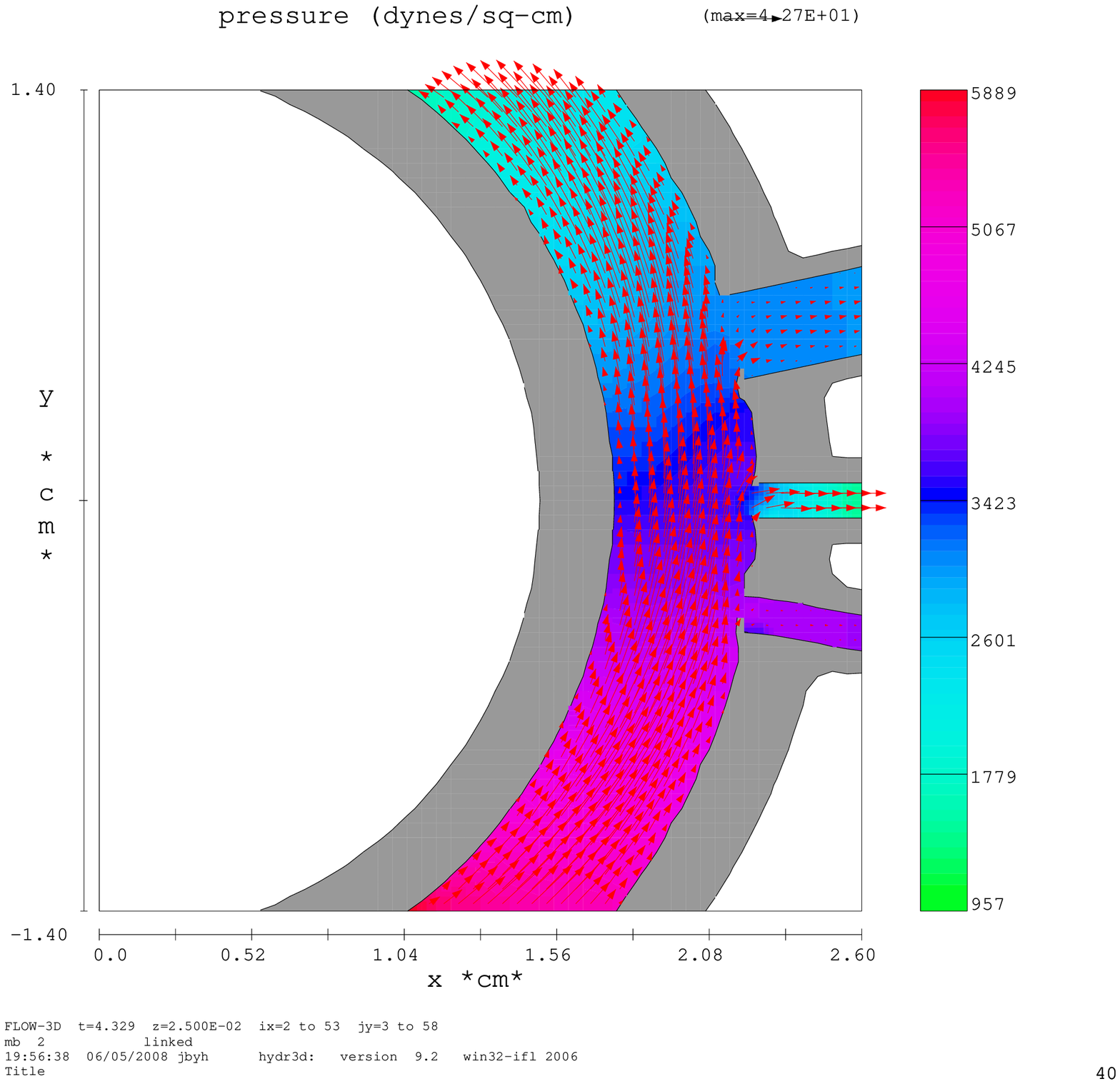}
\includegraphics[scale=1.0,width=25pc,height=25pc]{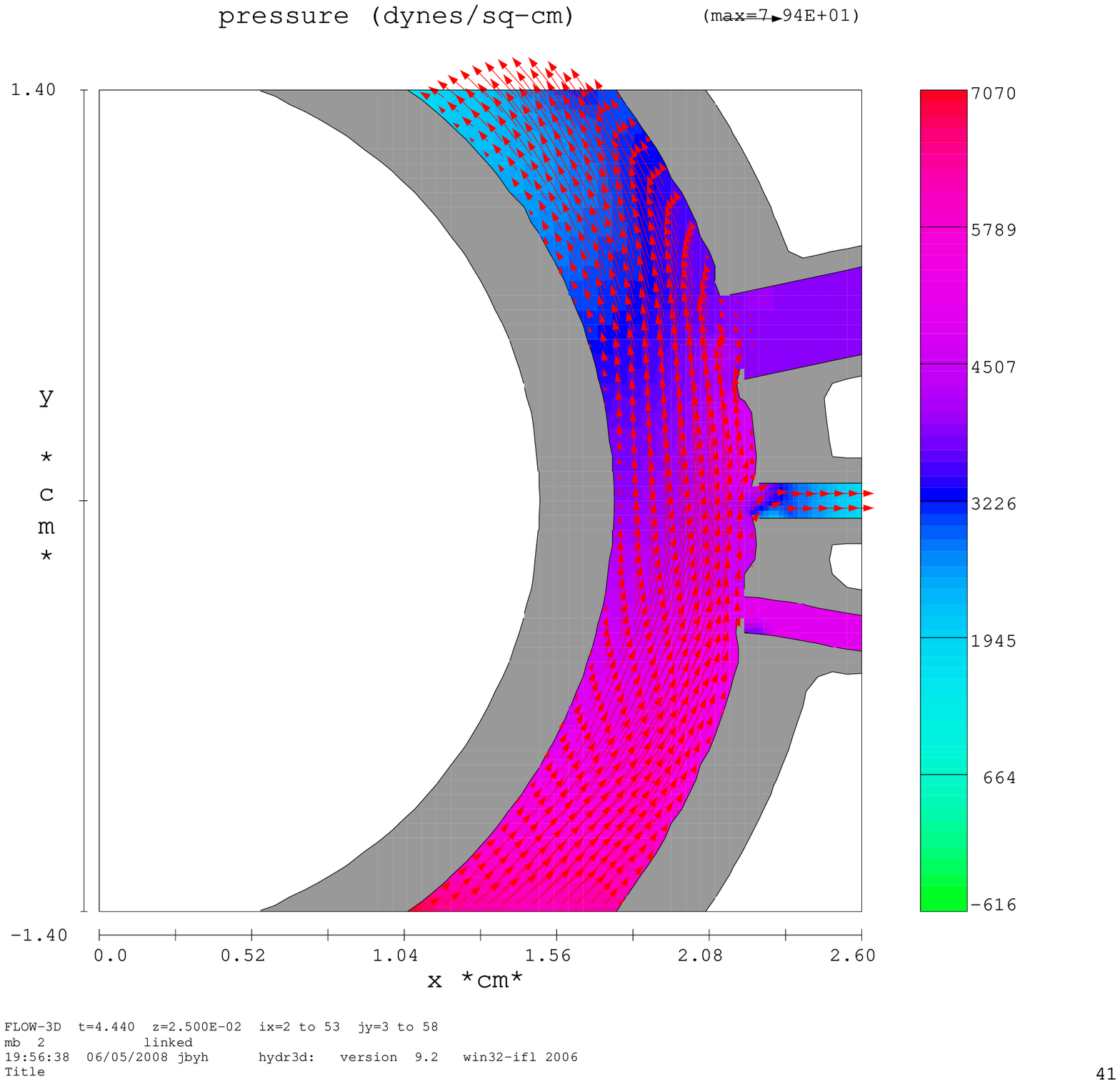}\\
\end{center}
\caption{These two plots represent in more detail
the region of the arch together with shaded results for the pressure distribution.
The bars on the right show the values.
These results are for two specific moments of the time 
$t_{40}$ = 4.329 sec and $t_{41}$ = 4.440 sec, where the pressure ranges
from 957 dynes/sp-cm to 5889 dynes/sq-cm (upper plot), and from -616 dynes/sq-cm
to 7070 dynes/sq-cm (lower plot).}
\label{fig:fig8}
\end{figure}
\end{document}